\documentclass[reprint,prl,preprintnumbers,amsmath,amssymb]{revtex4-1}
\usepackage{hyperref}
\usepackage{natbib}
\usepackage[dvips]{graphicx}
\usepackage{dcolumn}
\usepackage{bm}

\begin{document}
\bibliographystyle{unsrtnat}
\title{\large {Observation of a warped helical spin-texture in Bi$_2$Se$_3$ from circular dichroism angle-resolved photoemission spectroscopy}}

\vspace{0.5cm}

\author{Y. H. Wang$^{1,2}$}
\author{D. Hsieh$^{1}$}
\author{D. Pilon$^{1}$}
\author{L. Fu$^{2}$}
\author{D. R. Gardner$^{1}$}
\author{Y. S. Lee$^{1}$}
\author{N. Gedik$^{1}$}
\affiliation{$^1$Department of Physics, Massachusetts Institute of
Technology, Cambridge MA 02139, USA} \affiliation{$^2$Department of
Physics, Harvard University, Cambridge MA 02138, USA}

\vspace{0.5cm}

\begin{abstract}
A differential coupling of topological surface states to left-
versus right-circularly polarized light is the basis of many
opto-spintronics applications of topological insulators. Here we
report direct evidence of circular dichroism from the surface states
of Bi$_2$Se$_3$ using a laser-based time-of-flight angle-resolved
photoemission spectroscopy. By employing a novel sample rotational
analysis, we resolve unusual modulations in the circular dichroism
photoemission pattern as a function of both energy and momentum,
which perfectly mimic the predicted but hitherto un-observed
three-dimensional warped spin-texture of the surface states. By
developing a microscopic theory of photoemission from topological
surface states, we show that this correlation is a natural
consequence of spin-orbit coupling. These results suggest that our
technique may be a powerful probe of the spin-texture of spin-orbit
coupled materials in general.
\end{abstract}
\maketitle

Three-dimensional topological insulators
\cite{MooreReview,KaneReview,QiZhangReview} are an intensely
researched phase of matter owing to their unique spin-helical
metallic surfaces, where the spin direction is locked to the
wavevector and winds by $2\pi$ around the Fermi surface
\cite{Hsieh,ref26,Pan}. Spin-dependent absorption of circularly
polarized light has recently been predicted to generate highly
spin-polarized surface electrical currents whose direction can be
switched by the light helicity \cite{Raghu,Lu,hosur}. However, a
demonstration of spin-dependent differential absorption of left-
versus right-circularly polarized light [circular dichroism (CD)]
from the surface states (SS) has evaded conventional probes such as
transport and optics because of the combined need for
energy-momentum resolution and surface sensitivity.
\begin{figure}[t]
\includegraphics[scale=0.65]{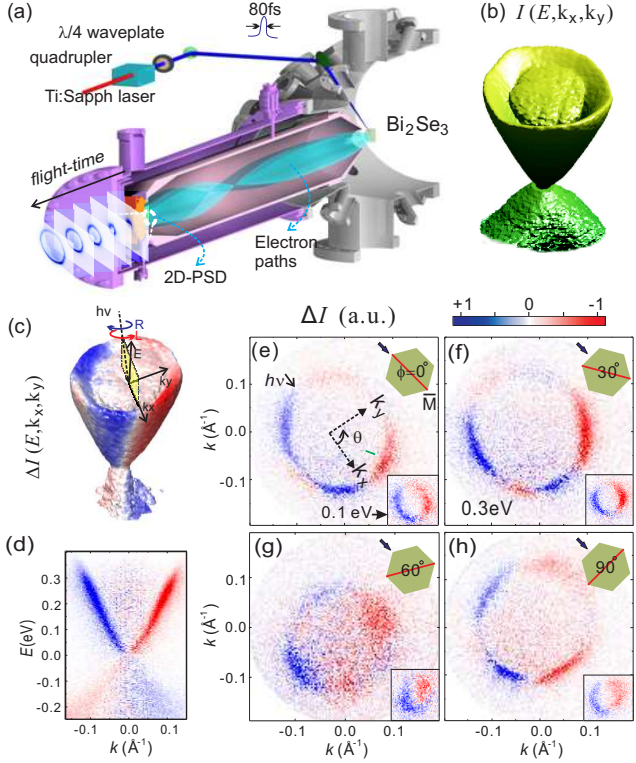}
\caption{(a) Schematic of a time-of-flight based angle-resolved
photoemission spectrometer. PSD: position-sensitive detector. (b) A
typical iso-intensity surface in $(E,k_x,k_y)$ space from
Bi$_2$Se$_3$ collected simultaneously using linearly polarized
photons. (c) Difference of TOF-ARPES data measured using right- and
left-circularly polarized light. $h\nu$ denotes the incident photon
direction. (d) $E-k$ cut through the CD data volume in (c) along the
angle denoted by the green dash in (e). (e-h) Constant energy slices
through (c) at the Fermi level and at 0.1eV above the DP (insets)
for sample rotation angles of $\phi=0^{\circ}, 30^{\circ},
60^{\circ}$ and $90^{\circ}$ respectively. Green hexagons represent
the Brillouin zone of Bi$_{2}$Se$_{3}$(111), red lines are the
mirror planes and the arrow denotes the photon incident direction.}
\label{fig:Fig1}
\end{figure}

In this Letter, we map the variation of the CD in the photoemission
intensity
\cite{Westphal,Schneider,Halilov,Frentzen,Zabolotnyy,Vyalikh} over
the full surface band structure of the topological insulator
Bi$_2$Se$_3$ \cite{Xia,ZhangNatPhy} using a laser-based
time-of-flight angle-resolved photoemission spectroscopy (TOF-ARPES)
technique. Our results show a strong CD from the SS in agreement
with a recent work on Cu$_x$Bi$_2$Se$_3$ \cite{Ishida}. However,
owing to our ultra-high polarization purity and our novel sample
rotational analysis, we uniquely resolve fine modulations in the CD
pattern as a function of energy and momentum that perfectly follow
the predicted \cite{FuHexWarp,ref29}, but so far un-detected
\cite{Pan}, three-dimensional spin-texture of Bi$_2$Se$_3$. By
developing a microscopic theory of photoemission from topological
SS, we show that such modulations naturally arise from the coupling
of the circularly polarized light to the surface spin-texture via
spin-orbit interactions. Excellent agreement between our results and
theory suggests that our technique is an efficient and sensitive
measure of all three components of the surface spin-texture.

The experimental setup is shown in Fig. 1(a). Single crystals of
Bi$_2$Se$_3$ are cleaved along their (111) surface at room
temperature under ultra-high-vacuum ($<6\times10^{-11}$ Torr).
Femtosecond laser pulses from a Ti:Sapph amplifier are frequency
quadrupled to 6.2eV and then circularly polarized to $>99\%$ purity
with a quarter wave-plate. Strain induced birefringence from the
vacuum windows and dichroism from all optical elements following the
quarter wave-plate, which can ruin the purity of circular
polarization, were carefully compensated in our experiment
\cite{SM}. Conventional hemispherical ARPES analyzers rely on the deflection of electron paths along one spatial dimension to measure electronic dispersion along one direction in momentum space. In contrast, our TOF-ARPES analyzer is developed for pulsed photon sources and resolves electron energies through their flight time from sample to detector, which enables measuring dispersion spectra $I(E, k_x,
k_y)$ simultaneously over two dimensional momentum space. Complete
spectra of Bi$_2$Se$_3$ (surface Dirac cone plus bulk conduction
band) obtained with TOF-ARPES is shown in Fig.1(b). This ability to
measure the entire bandstructure simultaneously eliminates
experimental geometry induced matrix element effects \cite{Hufner}
that may obscure the intrinsic CD pattern and allows for direct
comparison of CD-ARPES intensities across different phase space
points.

To investigate whether CD is exhibited by Bi$_2$Se$_3$ and if it has
bulk and/or surface state origin, we measure spectra like in
Fig.1(b) using left- and right-circularly polarized light and then
take their difference to obtain the CD spectrum $\Delta I(E, k_x,
k_y)$. As shown in Figs. 1(c)-(d), there is clear CD from the
surface Dirac cone of Bi$_{2}$Se$_{3}$ and not from the bulk
conduction band. The CD pattern exhibits perfectly odd reflection
symmetry about the photon scattering plane when it is parallel to
the $\bar{\Gamma} \bar{M}$ direction [Fig. 1(e)], which is expected
from crystal mirror symmetry. However this odd reflection symmetry
is lifted when the scattering plane is along a non mirror-symmetric
direction such as $\bar{\Gamma} \bar{K}$ [Fig. 1(f)], which is
evidence that the CD pattern is intrinsic to the sample.

Having shown that CD is intrinsic to the sample and present for the
SS, we first focus on its fine features in the low energy region
[$\pm0.1$eV relative to the Dirac point (DP)] where the Fermi
surface is circular. Fig. 1(d) shows that at a constant energy the
sign of CD is reversed for states of opposite momentum and is
reversed again across the DP. Furthermore, the CD pattern is
invariant under rotation ($\phi$) of the crystal mirror plane
relative to the scattering plane [Figs. 1(e)-(h) insets]. These
behaviors match the known in-plane spin texture in the low energy
region, which are tangential to the rotationally isotropic constant
energy contours \cite{Hsieh,ref26} and reverse chirality across the
DP.

The CD patterns change drastically away from the low energy region
($|E|>0.1$eV), where the Fermi surface evolves from being circular
to hexagonal in shape \cite{FuHexWarp,Kuroda}. Here we observe new
modulations in the magnitude and sign of CD as a function of the
angle $\theta$ around the constant energy contours [Figs.1(e)-(h)].
In particular, extrema along the $k_y$ axis (at $\theta=90^{\circ}$
and $270^{\circ}$) at low energy [Figs. 1(e)-(h) insets] become
nodes in the high energy $\phi=0^{\circ}$ spectrum [Fig. 1(e)].
Furthermore, in stark contrast to the low energy region, CD patterns
at high energy undergo dramatic changes as $\phi$ is varied. For
instance, the nodes along $k_y$ in the Fig. 1(e) spectrum become
extrema again in the $\phi=60^{\circ}$ spectrum [Fig. 1(g)] and only
repeats itself under $\phi=120^{\circ}$ rotations, consistent with
the symmetry of the underlying lattice. This is reminiscent of the
predicted $\theta$ dependence of the out-of-plane component of spin
$\langle S_z \rangle$ at high energies \cite{FuHexWarp,ref29}, which
goes from minimum to maximum under $\phi=60^{\circ}$ rotation. Such
behavior has been observed using Mott polarimetry only in
Bi$_2$Te$_3$ \cite{ref26} but not in Bi$_2$Se$_3$ \cite{Pan} owing
to the much larger value of $\langle S_z \rangle$ in the former.
Given that the overall CD patterns observed closely follow the
predicted spin-texture of the SS, we investigate their possible
connection by developing a standard model of photoemission from
topological SS.

The microscopic Hamiltonian for a system with spin-orbit coupling is
given by:
\begin{equation}
H= \frac{{\vec P}^2}{2m} + V(\vec{r}) + \frac{\hbar}{4m^2 c^2}
({\vec P} \times \vec \nabla V) \cdot  \vec s
\end{equation}
where $\vec P$ is momentum operator, $V(\vec{r})$ is the crystal
potential, and $\vec s$ is the electron spin. Coupling to an
electromagnetic field is obtained via ${\vec P} \rightarrow {\vec P}
- e {\vec A}$, where $\vec{A}$ is the photon vector potential, such
that to first order in $\vec{A}$:
\begin{equation}
H({\vec A}) = H - \vec{\cal P} \cdot {\vec A} \label{ha}
\end{equation}
where $\vec{\cal P}\equiv\frac{e}{m} {\vec P} - \frac{\hbar e}{4m^2
c^2} ({\vec \nabla} V \times {\vec s})$. The photoemission matrix
element between the initial and final states is given by
\begin{eqnarray} \label{m}
M(\vec{k},f) = \langle f_{\vec k} | \; \vec{\cal P} \cdot \vec{\cal
A}|\vec{k}\rangle
\end{eqnarray}
where ${\cal{\vec A}} \equiv \int d t {\vec A}(t) e^{i\omega t}$ is
the Fourier transform of $\vec{A}$ and $|f_{\vec k}\rangle$ is the
bulk final state. The initial state $|\vec{k}\rangle=u_{\vec{k}} |
\phi^i_+ \rangle  + v_{\vec{k}} | \phi^i_-\rangle$ is a linear
combination of two-fold degenerate pseudospin states $|\phi^i_\pm
\rangle$ at the DP that are eigenstates of total angular momentum
(orbital plus spin), which is widely used in standard $k\cdot p$
descriptions of topological SS
\cite{KaneReview,FuHexWarp,ZhangNatPhy,Basak}. Because of strong
spin-orbit coupling, only pseudospin is a good quantum number
\cite{FuHexWarp}. The coefficients $u_{\vec{k}}$ and $v_{\vec{k}}$
determine the expectation value of three pseudospin components: $
\langle S_x \rangle_{\vec k}= \hbar(u_{\vec k}^* v_{\vec k}
+v_{\vec k}^* u_{\vec k})$, $\langle S_y\rangle_{\vec
k}=\hbar(v_{\vec k}^* u_{\vec k} - i  u_{\vec k}^* v_{\vec k})$,
$\langle S_z\rangle_{\vec k} =\hbar(|u_{\vec k}|^2 -|v_{\vec
k}|^2)$. Importantly, because spin is directly proportional to
pseudospin \cite{FuHexWarp}, we refer to the two interchangeably.
For circularly polarized light incident onto the surface with
wavevector in the $xz$ plane, $\vec {A}(t)$ = $(A_x \sin \omega t ,
A_y \cos \omega t , A_z \sin\omega t )$ and $\cal{\vec A}$ =
$(-\mathrm{i} A_{x}, A_{y}, -\mathrm{i} A_{z})$. Straight-forward
application of time-reversal and crystal symmetries \cite{SM} yield
the following expression for the photoemission transition rate:
\begin{eqnarray}\label{m2}
I(\vec{k}) &=&  a^2 \left( |{\cal A}_x|^2 + |{\cal A}_y|^2 \right)  + b^2 | {\cal A}_z|^2 + a^2 {\rm Im} \left({\cal A}_x {\cal A}_y^* \right) \langle S_z\rangle_{\vec k} \nonumber \\
 &+& 2 a b  {\rm Im}[{\cal A}_x^* {\cal A}_z\langle S_y \rangle_{\vec k}- {\cal A}_y^* {\cal A}_z\langle S_x \rangle_{\vec k}]
\end{eqnarray}
where $a$ and $b$ are bandstructure dependent complex constants and
$Im$ refers to the imaginary part. Circular dichroism is obtained by
taking the difference of Eq.(\ref{m2}) with opposite photon helicity
($A_y\rightarrow-A_y$):
\begin{equation}
\label{diff} \Delta I = a^2 {\rm Im} ({\cal A}_x {\cal
A}_y^*)\langle S_z \rangle_{\vec k} - 4ab {\rm Im} ({\cal A}_z {\cal
A}_y^*) \langle S_x \rangle_{\vec k}
\end{equation}

Although the CD is a linear combination of $\langle S_x \rangle$ and
$\langle S_z\rangle$, these can be disentangled by applying symmetry
properties of Bi$_2$Se$_3$(111) as follows. Time-reversal symmetry
and three-fold rotational symmetry together dictate that $\langle
S_z \rangle$ flips sign upon a $\Delta\phi=60^{\circ}$ rotation
while $\langle S_x \rangle$ stays unchanged. Taking the difference
(sum) of CD patterns $\Delta\phi=60^{\circ}$ apart isolates $\langle
S_z \rangle$ ($\langle S_x \rangle$). The $\langle S_y \rangle$
component is trivially obtained by performing the procedure for
$\langle S_x\rangle$ under a $90^{\circ}$ sample rotation. Fig. 2
shows all three spin components measured using this method. Because
calculating $a$ and $b$ is beyond the scope of our work, we only
plot the relative and not absolute magnitude of the spins. We
immediately notice that all three components reverse sign across the
DP as expected and that $\langle S_x \rangle$ and $\langle S_y
\rangle$ are modulated with a similar periodicity. On the other hand
$\langle S_z \rangle$ exhibits a different periodicity at high
energies, which vanishes in the low energy region.

\begin{figure}[t]
\includegraphics[scale=0.65]{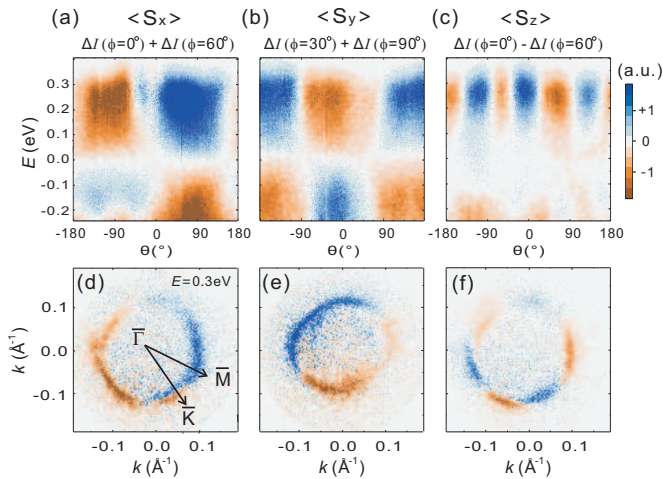}
\caption{(a) $x$, (b) $y$  and (c) $z$ components of the
spin-texture over the complete surface states obtained by summing or
subtracting $\Delta I$ data volumes at different $\phi$ (see text).
Color maps were created by integrating the data radially in k-space
over a $\pm0.015 \AA^{-1}$ window about the surface state contours
at each energy. Constant energy cuts at the Fermi level for each
spin component are shown directly below in (d) through (f).}
\label{fig:Fig2}
\end{figure}

\begin{figure}[t]
\includegraphics[scale=0.6]{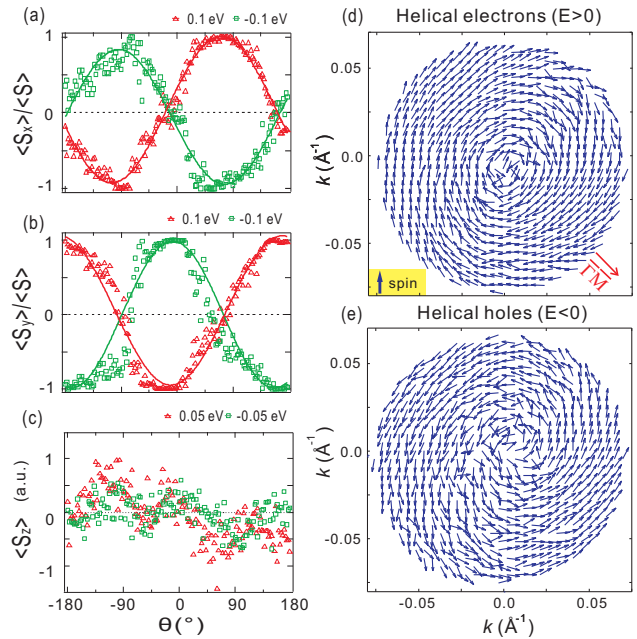}
\caption{The evolution of the (a) $S_x$, (b) $S_y$ and (c) $S_z$
components around constant energy contours in the vicinity of the
Dirac point, obtained by taking cuts through the data in Figs.
2(b)-(d). Curves are fits to $\sin(\theta)$ and $\cos(\theta)$
varying functions for $S_x$ and $S_y$ respectively. (d) and (e) show
the spin vectors (each normalized by its magnitude $|S|$) of all
positive and negative energy states within $\pm0.1$eV respectively
projected onto the $k_{x}-k_y$ plane.} \label{fig:Fig3}
\end{figure}
To quantitatively test the validity of our microscopic theory, we
first compare our extracted spin components with the ideal spin
texture in the low energy region that is experimentally well-known .
Constant low energy slices through the spin maps [Figs. 2(a)-(c)]
show that $\langle S_z \rangle$ exhibits negligible modulation
whereas $\langle S_x \rangle$ and $\langle S_y \rangle$ follow a
clear $\sin(\theta)$ dependence that reverses sign across the DP
[Figs. 3(a)-(c)]. By projecting the planar spin vectors of each
state in the low energy region of the Dirac cone onto the $x-y$
plane, we obtain a vector field that exactly matches an ideal
helical spin-texture with opposite chirality for electrons and holes
[Figs. 3(d) and (e)]. We note that while the relative orientation of
spins is directly measured in our experiment, the absolute sense of
chirality was set by matching our data to Mott polarimetry results
at a single energy-momentum point \cite{Pan}, although it can also
be independently obtained by calculating $a$ and $b$. The excellent
agreement between our measured low energy CD maps and theoretical
\cite{FuHexWarp} and Mott polarimetry results \cite{Hsieh,ref26,Pan}
strongly suggests that CD-ARPES is a sensitive measure of the
topological SS spin-texture.

Next we compare our CD spectra to our model in the high energy
region where the spin-texture is predicted to depart from ideal
planar helical form but is so far un-observed in Bi$_{2}$Se$_{3}$
using conventional Mott polarimetry based ARPES \cite{Pan}. At high
energies where the Fermi surface acquires a hexagonal shape, $k\cdot
p$ theory predicts that the spin-texture should develop a finite
out-of-plane component $\langle S_z \rangle$ that is modulated with
a $\sin(3\theta)$ periodicity, with maximally upward and downward
spin canting along $\bar{\Gamma} \bar{K}$ directions
\cite{FuHexWarp}. Fig. 4(a) shows a constant energy slice at 0.3eV
through the spin map of Fig. 2(c), where there is a clear
$\sin(3\theta)$ dependence of $\langle S_z \rangle$ that is maximal
along $\bar{\Gamma} \bar{K}$ and zero along $\bar{\Gamma} \bar{M}$,
in perfect accord with both theory \cite{FuHexWarp} and experiment
\cite{ref26} on Bi$_2$Te$_3$. To further test whether our CD-ARPES
data is indeed sensitive to spin, we compare our measured energy
dependence of out-of-plane spin component along $\bar{\Gamma}
\bar{K}$ ($\langle S_z \rangle^{0} (E)$) with its unique functional
form predicted by $k\cdot p$ theory \cite{FuHexWarp}:
\begin{equation}\label{szfit}
\langle S_z \rangle^{0} (E)=1/\sqrt{1+[k(E)\beta]^{-4}}
\end{equation}
where $\beta$ is the only free parameter. Eq.(\ref{szfit}) provides
an excellent fit for our data [Fig. 4(c)] and the fitted value of
$\beta$, which parameterizes the degree of hexagonal distortion of
the Fermi surface, is quantitatively consistent with the parameters
extracted from fitting the Fermi surface contour in Fig. 4(c) inset
\cite{SM}. The accurate extraction of all three spin components over
the entire Dirac cone from our CD-ARPES data [Fig. 4(d)] is a
testament to the validity and sensitivity of our method.

\begin{figure}[t]
\includegraphics[scale=0.68]{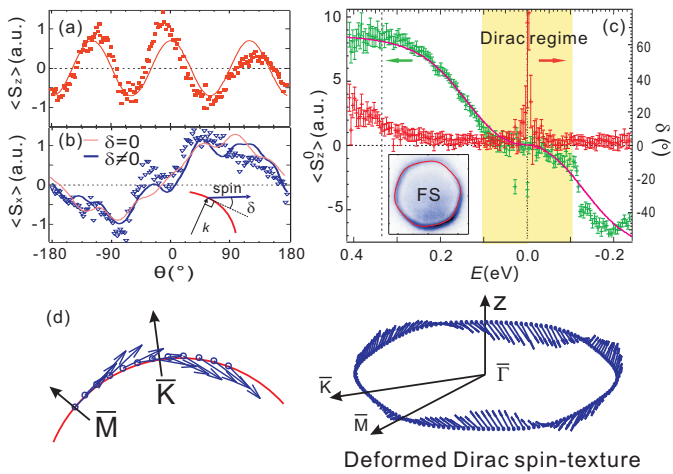}
\caption{ Evolution of (a) $S_z$ and (b) $S_x$ around the Fermi
surface. Red line in (a) is a fit to $k\cdot p$ theory. The blue and
pink lines in (b) are fits to phenomenological models \cite{SM} that
respectively do and do not account for an in-plane angular deviation
($\delta$) away from $\vec{S}\perp\vec{k}$ locking behavior (inset).
(c) The energy dependence of the amplitude of $S_z$ modulation
(green symbols), its fit (magenta line) and $\delta$ (red symbols).
Inset shows the spin-integrated Fermi surface together with a fit to
$k\cdot p$ theory. (d) Spin-orientation around the hexagonally
deformed Fermi surface constructed from CD-ARPES data. 3D view shows
spins plotted around Fermi contour and top-down view shows spins
projected onto a segment of the unit circle in momentum space for
ease of visualizing $\delta$.} \label{fig:Fig4}
\end{figure}

While our data produce all features predicted by $k\cdot p$ theory
\cite{FuHexWarp}, the in-plane spin component at high energies
also exhibits unpredicted kinks in its $\theta$ dependence [Fig.
4(b)]. One possible origin of these kinks is that the in-plane
spin magnitude is modulated purely by the out-of-plane spin
canting. However a fit to the data using this assumption [Fig.
4(b) pink line] yields kinks that are too weak and fails to
reproduce the nodes at $\theta=0^{\circ}$ even when allowing for
unrealistically large ($\sim 45^{\circ}$) canting angles
\cite{SM}. An alternative explanation is that the in-plane spin
direction deviates from perpendicular spin-momentum locking [inset
Fig. 4(b)] periodically in $\theta$. A fit assuming periodic
angular deviations of amplitude $\delta$ [Fig. 4(b) blue line]
better reproduces the data. The in-plane component constructed
from the data alone [Fig. 4(d)] is consistent with the latter
interpretation and shows that the deviation is zero along
$\bar{\Gamma} \bar{M}$ and $\bar{\Gamma} \bar{K}$ and is maximum
in between the two. A plot of the amplitude of $\delta$ versus
energy [Fig. 4(c)] shows that it only develops beyond $\sim0.3eV$
from the DP. The spike exactly at the DP is consistent with spin
degeneracy at the DP. This in-plane spin canting behavior has been
shown by first-principles calculations as well as $k\cdot p$
expansions up to fifth order on Bi$_{2}$Te$_{3}$ \cite{Basak}.
However, although such effects have been observed in Rashba spin
split metals \cite{Meier}, they have never been observed in any
topological insulators.

In conclusion, we have directly observed circular dichroism from the
surface states of Bi$_{2}$Se$_{3}$. By combining a TOF-ARPES
technique with ultra-pure circular photon polarization and a novel
sample rotation analysis, we resolve unusual modulations in the
circular dichroism photoemission pattern as a function of both
energy and momentum for the first time, which closely follow the
predicted three-dimensional spin-texture. A direct connection
between CD-ARPES and spin-orbit induced spin-textures is established
through our microscopic theory of photoemission. Our results open
the possibility to generate highly-polarized spin currents with
circularly polarized light, which may be detected through transport
or optical means \cite{HsiehSHG}. The efficiency and high spin
sensitivity of our technique suggest that CD-ARPES may be used as a
vectorial spin mapping tool to detect small deviations from a $\pi$
Berry's phase in magnetically doped topological surface states, or
to study spin-orbit coupled materials in general.

Towards completion of this manuscript, we became aware of three
related concurrent and independent works \cite{Ishida,Park}. None,
however, resolve the fine modulations in the CD-ARPES patterns that
are integral to understanding the connection to spin, which is the
main focus of this work.

This research is supported by Department of Energy award number
DE-FG02-08ER46521, Army Research Office (ARO-DURIP) award number
W911NF-09-1-0170 (ARTOF spectrometer) and in part by the MRSEC
Program of the National Science Foundation under award number DMR
- 0819762 (partial support for YHW). Correspondence and requests
for materials should be addressed to N.G. (gedik@mit.edu).

\vspace{1cm}
\newpage

\end{document}